\documentclass[pdftex,12pt]{JHEP3}
\usepackage{amsmath,amsfonts,amsbsy,latexsym,amssymb,amscd,amstext}
\usepackage[pdftex]{epsfig}
\usepackage{graphicx}
\usepackage{caption}

\pdfoutput=1

\def\be{\begin{equation}}
\def\ee{\end{equation}}
\def\bea{\begin{eqnarray}}
\def\eea{\end{eqnarray}}
\def\pd{\partial}
\def\a{\alpha}
\def\b{\beta}
\def\g{\gamma}
\def\d{\delta}
\def\m{\mu}
\def\n{\nu}
\def\t{\tau}

\def\l{\lambda}

\def\r{\rho}
\def\s{\sigma}
\def\e{\epsilon}

\def\bi{\begin{itemize}}
\def\ei{\end{itemize}}

\def\bp{\bar{\phi}}

\date{May 25th, 2008} \preprint{DFT-UAM-10-05\\IFT-UAM/CSIC-10-24}
\title{Comments on the vacuum energy decay} \author{Enrique \'Alvarez, Roberto Vidal \\  Instituto de F\'{\i}sica Te\'orica
UAM/CSIC and Departamento de F\'{\i}sica Te\'orica \\ Universidad
Aut\'onoma de Madrid, E-28049--Madrid, Spain \\ E-mail: \email{enrique.alvarez@uam.es}, \email{jroberto.vidal@uam.es}}
\abstract{The {\em instability of vacuum energy} in de Sitter space as discussed recently by Polyakov is argued to be a generic feature when external gravitational fields are present. It is related to the existence of {\em forbidden} (by momentum conservation)  decays  derived in some detail by Bros, Epstein and Moschella. Some calculations are discussed in a conformally invariant setting, and a flat space model is presented.

}
\begin{document}
\section{Introduction}

The understanding of the cosmological constant and its relationship with the vacuum energy is one of the most important problems in theoretical physics as of today. A recent contribution to it is Polyakov's \cite{Polyakov} claim that {\em vacuum energy is unstable.} This instability is supposed to be different from previous proposals in the sense that this is due to particle interactions in a de Sitter background, and consequently is only indirectly of gravitational origin.
\par
To be specific, what is asserted is that interactions draw a generic instability as an imaginary part of the free energy of the quantum fields as computed in a de Sitter background.
\par
As a matter of fact Polyakov's  is a  three-loop effect  and, as such, lies outside of our previous analysis \cite{Alvarez}, which was restricted to one-loop order. Our purpose in the present paper is to briefly comment  on some physical aspects of this effect. 
\par
 To begin with, it is worth pointing out the following. In an interesting series of papers, Bros, Epstein and Moschella \cite{Bros} following early work \footnote{In the book by Birrell and Davies \cite{Birrell} some earlier references can be found.
}
in \cite{Nachtmann} and \cite{Myhrvold}, 
have shown that one particle decays in $ \phi^3$ or $\phi^4$ theories are not forbidden kinematically in de Sitter space. Representing by $h(x)$ the scalar field, such decays imply a nonvanishing width
\[
\Gamma\left(h\rightarrow h h\right)\text{ or else }\Gamma\left(h\rightarrow h h h\right)
\]

This is in sharp contrast with the situation in flat space, where momentum conservation forbids them.  The reason for that is the lack of translational invariance (there is no abelian translation subgroup of the de Sitter group; pseudotranslations do not commute), so that two-point functions are not necessarily functions of the difference between spacetime coordinates of the two points, which is the root of global momentum conservation in any physical process. In fact this effect is common to any quantum field theory in a nontrivial gravitational background.

\par
Once momentum conservation is not working, nothing forbids the vacuum decay in to physical particles, which essentially related to the effect pointed out by Polyakov. Assuming, as we do, {\em crossing symmetry}, the preceding channels are related to the vacuum decay in the tree approximation
\[
\Gamma\left(0\rightarrow h h h\right) \text{ or else } \Gamma\left(0\rightarrow h h h h\right) 
\]
\par
Once there is a nonvanishing  amplitude for this sort of decay into several particles, it seems plain that the inverse reaction is much less likely, so that there is an enhanced production until the particle density $n$ is so high that
\[
n^{1/3}\sim \Gamma
\]
at which point detailed balance should establish itself and the particle production growing stops.

In a curved space the concept of particle depends on the observer  through the definition of positive frequencies associated to a complete system of solutions of the classical field equations. For example, if $f^1_i(x)$ and $f^2_i(x)$ are two such systems appropriate in the far past and in the far future of the process of interest, the \emph{free} quantum field can be expanded as

\[
\phi(x)=\sum_n\left(a_n f^1_n(x)+a^\dagger_n f^{1 *}_i(x)\right)=\sum_n\left(b_n f^2_n(x)+b^\dagger_n f^{2 *}_n(x)\right)
\]
This defines the operators $a_n$ and $b_n$. This is also true, in particular, for the asymptotic fields, $\phi_\textrm{in}$ and $\phi_\textrm{out}$. Now four different vacua can be defined as usual in an S-matrix approach
\begin{eqnarray}
a^\textrm{in}_n|0^1_\textrm{in}\rangle=0 && a^\textrm{out}_n|0^1_\textrm{out}\rangle=0 \nonumber\\
b^\textrm{in}_n |0^2_\textrm{in}\rangle=0 && b^\textrm{out}_n |0^2_\textrm{out}\rangle=0\nonumber
\end{eqnarray}

The two sets of modes are related by a Bogoliubov transformation, so that
 a given vacuum contains particles as defined with respect to a different vacuum (that is, using a different definition of positive frequency).
When interactions are taken into account {\em even with the same definition of positive frequency} the vacuum is not stable 
 \[
|\langle 0_\textrm{out}^{1,2}| 0_\textrm{in}^{1,2}\rangle|\neq 1
\]

In practice both effects (that is, particle creation due to the external gravitational field at zero coupling as reported, for example, in  \cite{Mottola}, and the effects of the interaction) compete, and in order to separate them one has to study specific channels as well as their dependence on the coupling.\\
\par

The \emph{physical} matrix elements to calculate in a $\alpha\to\beta$ are referred to the appropriate notion of particles $\vphantom{a}_\textrm{out}\langle\beta^2|\alpha^1\rangle_\textrm{in}$, i.e. choosing the modes corresponding to the (asymptotic) past and future zones where the field behaves as a free field. Even when S-matrix elements are not defined {\em sensu stricto} (such as in de Sitter space) transition amplitudes for finite time intervals can still be computed using Feynman's rules. Our point of view is similar to the one in \cite{Einhorn} in that we assume that  enough of the analytical scheme of flat space quantum field theory survives to justify the formal use of the interaction representation and related path integral techniques.\\

\par

In order to separate the true interaction effects from the creation of particles due to the gravitational field, we need to focus in the $\vphantom{a}_\textrm{out}\langle\beta|\alpha\rangle_\textrm{in}$ matrix elements, referred to a common notion of vacuum, which are related by a Bogoliubov transformation to the previous ones.\\

For this matrix elements we can define the $S$-matrix as $S^\dagger|\alpha\rangle_\textrm{in}=|\beta\rangle_\textrm{out}$, which has the familiar interaction representation form 
\[
S=T\exp\{-i\int H_I(\phi)\}
\]
\par

In any $S$-matrix perturbative framework, unitarity precisely relates the imaginary part of the vacuum diagrams to creation and absorption of physical particles from the vacuum,
\[
S\equiv 1+i \mathcal T
\]
Unitarity means that for any couple of states $|a\rangle$ and $|b\rangle$, and any closure relation
\[
\sum|n\rangle\langle n|=1
\]
the following is true
\[
\langle a|b\rangle=\langle a|S S^\dagger|b\rangle=\langle a|\left(1+i \mathcal T\right)\left(1-i\mathcal T^\dagger\right)|b\rangle
\]
so that
\[
\langle a|i\left(\mathcal T-\mathcal T^\dagger\right)|b\rangle=-\langle a|\mathcal T \mathcal T^\dagger|b\rangle=-\sum_n \langle a|\mathcal T|n\rangle\langle n|\mathcal T^\dagger|b\rangle
\]

In a $\lambda\phi^4$ theory, to second order we have for the vacuum-to-vacuum amplitude:
\[
\langle 0|i\left(\mathcal T^{(2)}-\mathcal T^{(2)\dagger}\right)|0\rangle=-\sum_n \langle 0|\mathcal T^{(1)}|n\rangle \cdot\vphantom{a}\langle n|\mathcal T^{(1)\dagger}|0\rangle
\]
where we do not specify the appropriate asymptotic limit (i.e. \emph{in} or \emph{out}), since the obtained result is identical
\begin{equation}\label{im}
2\text{Im}\,\langle 0|\mathcal T^{(2)}|0\rangle=\sum_n |\langle 0|\mathcal T^{(1)}|n\rangle|^2 
\end{equation}

Up to second order, the $\mathcal T$ matrix for a $\lambda\phi^4$ theory is
\begin{align}\label{Tup2}
	S&=T\exp\left\{-i\frac\lambda{4!}\int\,dx\,:\phi(x)^4:\right\}\nonumber\\
\mathcal T&=-\frac\lambda{4!}\int\,dx:\phi(x)^4:+i\frac{\lambda^2}{2\cdot4!^2}\int\,dxdy\,T(:\phi(x)^4:\times:\phi(y)^4:)+\ldots
\end{align}
so the previous relation gives\footnote{We ignore the tadpole contributions that disappear by considering normal order in the interaction term.}

\begin{equation}\label{check}
\text{Re}\int dxdy\,G(x,y)^4=\int dx dy \,W(x,y)^4
\end{equation}
where $G(x,y)\equiv\langle0| T\phi(x)\phi(y)|0\rangle$ is the ``Feynman'' propagator and $W(x,y)\equiv\langle0|\phi(x)\phi(y)|0\rangle$ the Wightman function.

\section{Vacuum decay}\label{vacdecay}
  In Poincar\'e coordinates the metric of de Sitter space reads
\[
ds^2=\left({l\over u}\right)^{2}(du^2-d\mathbf x^2)
\]

A conformally coupled scalar field is massless and the value of the curvature coupling is $\xi=\frac14\frac{n-2}{n-1}$, where $n$ is the dimension of spacetime. In de Sitter space, where the curvature is constant, $R={n(n-1)\over l^2}$, this is equivalent to a minimally coupled ($\xi=0$) scalar field with mass $m^2=\frac14\frac{n-2}{n-1} R={n(n-2)\over 4l^2}$.
\par
 This mass is in the complementary series of the de Sitter group $SO(n,1)$, with a $i\mu=\frac12$ parameter ($m^2 l^2=\mu^2+{(n-1)^2\over4}$) , and the functional form of their two-point functions (without entering into the $i\e$ prescriptions for the time being)  is particularly simple 
 
\[
\frac{\Gamma\left(\frac n2-1\right)}{l^{n-2}(4\pi)^\frac n2} F\left(\frac n2,\frac n2-1,\frac n2;\frac{1+z}2\right)= \frac{\Gamma\left(\frac n2-1\right)}{2(2\pi)^\frac n2l^{n-2}}(1-z)^{1-\frac n2}
\] 
 \par
 Let us concentrate in this simplest example for the time being. In the coordinates we are using
\be
1-z={-\left(u-u^\prime\right)^2+\left(\vec{x}-\vec{x}^\prime\right)^2\over 2 u u^\prime}
\ee
 
Exploiting the conformal invariance of the setup \footnote{Broken by the interactions, however.}, we can expand the free field expansion in term of the $n$-dimensional modes
\[
f_\mathbf{k}(u,\mathbf x)={ u^{n/2 -1}\over (2\pi)^\frac{n-1}2 l^{n/2 -1}\sqrt{2k}}e^{-iku}e^{i\mathbf{kx}}
\]
which are equivalent to choosing the Euclidean or Bunch-Davies vacuum. 

\vspace{1cm}

It is a fact that the closure relation for the complete set of solutions gives a particular solution of the homogeneous equation (id est, without the delta function source) namely
\begin{align}\label{wightman}
\int d^{n-1}\mathbf k\, f_\mathbf{k}(u,\mathbf x)\,f^*_\mathbf{k}(u^\prime,\mathbf x^\prime)&=\frac{(uu')^{n-2}}{l^{n-2}}\int\frac{d^{n-1}\mathbf k}{(2\pi)^{n-1}2k}e^{-ik(u-u')}e^{i\mathbf k(\mathbf x-\mathbf x')}=\nonumber\\
&=\frac{\Gamma\left(\frac n2-1\right)}{2(2\pi)^\frac n2 l^{n-2}}\left(\frac{2uu'}{(i(u-u'-i\epsilon))^2+r^2}\right)^{\frac n2-1}
\end{align}
where the integrals can be done with
\[
\int_{-1}^1 dy (1-y^2)^\frac{n-4}2e^{i a y}=\sqrt\pi\left(\frac 2a\right)^\frac{n-3}2\Gamma\left(\frac n2-1\right)J_\frac{n-3}2(a)
\]
\[
\int_0^\infty dx J_\frac{n-3}2(\beta x)e^{-\alpha x} x^\frac{n-3}2=\frac{(2\beta)^\frac{n-3}2\Gamma\left(\frac n2-1\right)}{\sqrt\pi(\alpha^2+\beta^2)^{\frac n2-1}}
\]
and the last one needs an small real part for $\alpha$. This closure relation appears in the derivation of (\ref{check}), and this particular homogeneous solution is precisely the Wightman function. It is plain in (\ref{wightman}) that the $i\epsilon$ prescription depends on the time ordering of the arguments. The unitarity relations rely on this subtle difference with the Feynman propagator. \\


In the $\lambda \phi^4$ theory, the bubble diagram $G(x,y)^4$ can be computed for the conformally coupled case in 4 dimensions. The propagator is $\displaystyle\frac{1}{l^2}\frac{1}{z-1-i\epsilon}$ where $z$ is the geodesic distance, times a constant we will ignore. Given that de Sitter space is homogeneous, the diagram is proportional to the infinite spacetime volume $V_4\equiv\int d^4y\sqrt{\|g\|}$:
\[
M_{0\to0}=i\frac{\lambda^2}{2\cdot4!l^8} V_4\int d^4 x\sqrt{\|g\|}\frac{1}{(z(x,x_0)-1-i\epsilon)^4}
\]
and we can choose $x_0$ as an arbitrary point. Choosing $x_0$ as the north pole $N=(0,l,0,\ldots,0)$, we get
\[
M_{0\to0}=i\frac{\lambda^2}{2\cdot4!l^8} V_4\int d^4 x\sqrt{\|g\|}\frac{1}{(z(x,x_0)-1-i\epsilon)^4}=i\frac{\lambda^2(4\pi)}{2\cdot4!l^4} V_4\int dt\,d\theta \frac{\cosh^3 t\sin^2\theta}{(\cosh t\cos\theta-1-i\epsilon)^4}
\]

\begin{figure}[b]
\begin{center}
	\includegraphics[scale=0.50]{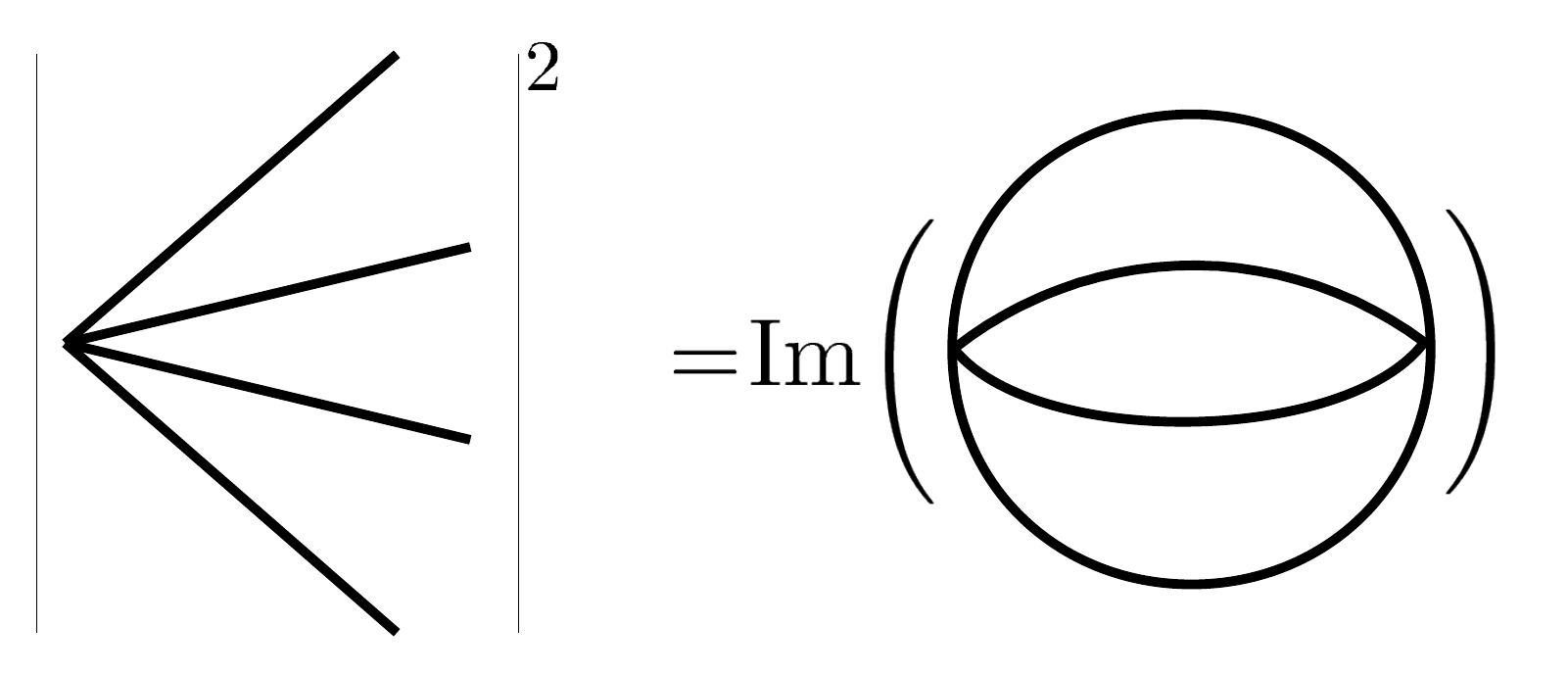}
	\caption{Unitarity relation in the vacuum-to-vacuum amplitude to second order.}
\end{center}
\end{figure}
To compute the imaginary part, we can use the well-known formula \cite{Gel'fand}
\[
{1\over (x-i0)^4}={1\over x^4}-i{\pi\over 6}\d^{(\text{iii})}(x)
\]
It follows that
\bea
\text{Im} M_{0\rightarrow 0}&&=\frac{\lambda^2(4\pi)}{2\cdot4!l^4}V_4\int dt\,d\theta\,\frac{\cosh^3 t\sin^2\theta}{(\cosh t\cos\theta-1)^4}=\nonumber\\
&&=\frac{\lambda^2(4\pi)}{2\cdot4!l^4}V_4 P\int {dx\over x^4} d\theta {\sin^2\,\theta\over \cos\,^3\theta}\frac{(1+x)^3\text{sgn}(x,\theta)}{\sqrt{(1+x)^2-\cos^2\theta}}
\eea
where the principal part $P$ regularizes the divergence at $x=0$ and $\text{sgn}(x,\theta)$ is zero except in the case $(1+x)\cos\theta>0$. It seems clear that the angular integral diverges owing to the behavior of the integrand in a neighborhood of $\theta={\pi\over 2}$. \\

This is a very strange result indeed, because as we shall see in a moment, this is equivalent to a corresponding divergence in the vacuum decay amplitude {\em in the tree approximation}, which has been reported \cite{Higuchi}  to be finite in the literature. This interesting calculation is reviewed in the Appendix.
\par
The result we get is
\be\label{int}
M_{0\rightarrow 0}=i\frac{\lambda^2(4\pi)}{2\cdot4!l^4}V_4\lim_{\e\rightarrow 0^+}\textrm{sgn}(\epsilon)\left(\frac{i\pi}{6\epsilon^2}+\frac\pi{6\epsilon}+\frac{i\pi}8+\ldots\right)\equiv i \frac{\lambda^2(4\pi)}{2\cdot4!l^4}V_4\lim_{\epsilon\to0^+}\cdot I(\epsilon)
\ee

This expression is divergent in the limit when $\e\rightarrow 0^+$, and it has got a finite part in this regularization of sorts we are using. Of course we are aware that only the limit when $\e\rightarrow 0^+$ has got any physical sense, so that the divergent parts are regularization dependent and just have to be renormalized away.
\par


Let us now turn our attention to the computation of the vacuum decay rate. We use again the de Sitter homogeneity as we did for the vacuum bubble\footnote{The Wightman functions are not fully de Sitter invariant (not under time reversal). However this is enough to repeat the procedure.}. The corresponding integral contains Wightman functions instead of propagators
\begin{align}
&|M_{0\to4}|^2=\frac{\l^2 V_4}{4!} \int d^4x \sqrt{\|g\|} \prod_{i=1}^{4} d\mathbf k_i f_{\mathbf{k}_i}(x)f^*_{\mathbf{k}_i}(x_0)=\nonumber\\
&=\frac{\lambda^2}{4!l^8}V_4\int d^4x\sqrt{\|g\|}\,W(x,x_0)^4=\frac{\lambda^2}{4!l^8}V_4\Big(\int_{X^0>0} d^4x\sqrt{\|g\|}\,G(x,x_0)^4+\nonumber\\
&+\int_{X^0<0} d^4x\sqrt{\|g\|}\,(G(x,x_0)^*)^4\Big)=\frac{\lambda^2(4\pi)}{4!l^4}V_4\left(\frac{I(\epsilon)}2+\frac{I(-\epsilon)}2\right)=\frac{\lambda^2(4\pi)}{4!l^4}V_4\cdot\frac\pi{6|\epsilon|}
\end{align} 
where we are again ignoring the concrete normalization factors of the two-point functions; and we have used that the integral of the propagator over \emph{half} of the full hyperboloid (over $t\in[0,\infty)$) is precisely the integral in (\ref{int}), $I(\epsilon)$, divided by two.

The vacuum decay  matches exactly the imaginary part of the vacuum energy in the regularized theory (as in (\ref{im})), which is  more than is necessary for unitarity, which does not necessarily hold for the regularized theory (id est, before taking the limit $\e\rightarrow 0^+$ in our case).
 \par
  It is worth remarking that the total {\em vacuum decay rate} per unit volume and unit time interval at tree level  appears to be divergent in the $\e\rightarrow 0$ limit. Given the fact that the corresponding tree level amplitude is indeed finite, the divergence is entirely due to the integration over the {\em phase space}. We are not aware of any such divergences, except the ones associated to bremsstrahlung corrections in the external legs \cite{DT}.  It would be interesting to study what happens in the non-conformal case as well as to investigate whether this divergences are related to the ones found in \cite{Akhmedov:2009ta} in a classical context.
  \par
  In the Appendix we discuss a different way of computing the vacuum decay rate, in a form that results proportional to the covariant four-volume in a way consistent with Fermi's Golden Rule; there seems to be an ambiguity as to whether the Rule can be applied in arbitrary coordinates.
  
  \par
   Indeed, in \cite{Polyakov} some arguments are given for finiteness (after substraction) of the massive contribution to the imaginary part of the effective action, and in \cite{Bros}\cite{Myhrvold} it is similarly argued for the finiteness of the {\em forbidden decay width}, which is, as is has been already argued for, a closely related quantity through crossing symmetry.


The result above is quite general, as these unitarity relations are ``built-in'' in the $S$ matrix formalism. In an $\lambda \phi^n$ theory with an arbitrary mass, the relevant identity to second order is
\begin{equation}
\text{Re}\int\,dxdy\,G(x,y)^n=	\int\,dxdy\,W(x,y)^n
\end{equation}
Using again the homogeneity of de Sitter, we obtain
\begin{equation}
\text{Re}\int\,dx\,G(x,x_0)^n=	\int\,dx\,W(x,x_0)^n
\end{equation}
and the relationship between the propagator and the Wightman function $G(x,y)=T(W(x,y))$ allows us to decompose these integrals
\begin{equation}
\int\,dx\,G(x,x_0)^n=I^++I^-\textrm{ , }\int\,dx\,W(x,x_0)^n=I^++(I^-)^*
\end{equation}
where we separate de Sitter space in two regions, future and past of $x_0$, and those are their respective contributions
\footnote{Actually the argument $x$ can be in the future, the past or be causally disconnected from $x_0$. We are splitting artificially the spatial region in two pieces (respect to some time parameter) and including them in the two causal contributions. This procedure is correct, because both the propagator and the Wightman function have the same real value in that region.}. So now our identity looks
\begin{equation}
\text{Im}{I^+}=\text{Im}{I^-}
\end{equation}

The decomposition of the Feynman propagator is 
\[
G(x,y)=\frac12(G^{(1)}(x,y)+i\sigma(x,y) D(x,y)),
\]
 where the Hadamard symmetric function $G^{(1)}$  and the Pauli-Jordan commutator function $D$ are real, $\sigma(x,y)$ is the sign of the time ordering of $x$ and $y$, and $D$ is antisymmetric and zero for causally disconnected points, i.e. has the form $D(x,y)=\sigma(x,y)\Delta(x,y)$ with $\Delta(x,y)$ symmetric.\\

The imaginary part of the $n$th power of the propagator is then proportional to $i\Delta(x,y)$ times asymmetric function (built with odd powers of $G^{(1)}$ and even powers of $D$), in such a way that the integrand depends only on $z(x,x_0)$, and not on the sign of the time ordering, so the transformation $X^0\to-X^0$ leaves unaltered the value of the integral. This proves the unitarity relation.

There is a formulation of unitarity (the {\em largest time equation})\footnote{To be precise, the statement is that in flat space the largest time equation can be shown to imply unitarity of the S-matrix.} which acts of position space Feynman diagrams themselves and, as such is most suitable for application in curved spacetimes. This set up is due to Veltman \cite{Veltman}, and asserts that
for any diagram $F\left(x_1,\ldots,x_n\right)$,
\[
2 \text{Re}\, F\left(x_1,\ldots,x_n\right)=-\sum_{cuttings}F\left(x_1,\ldots,x_n\right)
\]
This formula stems from a representation of the Feynman propagator as 

\[
\Delta_F\left(x,y\right)=\theta(t)\Delta_F^+(x,y)+\theta(-t)\Delta_F^-(x,y)
\]
such that
\[
\left(\Delta_F^+(x,y)\right)^*=\Delta_F^-(x,y)
\]
Those conditions are fulfilled in our case. It is always possible to argue that all we are doing is to check the largest time equation, without necessarily committing ourselves to the thorny \cite{Einhornl} issue of unitarity in de Sitter space. 
\par
In  reference \cite{Bros} it is stated that the quantitative importance of the  {\em forbidden} vacuum decays for massive particles as well as {\em small} values of the cosmological constant  is
\[
e^{-l |\Delta m|}
\]
where $\Delta m$ is the violation of energy conservation in the decay process. In fact, $m l =1$  corresponds roughly to the {\em quintessence} scale $ H_0\sim 10^{-33}\, eV$ at the present period of cosmic evolution.

\section{An imaginary part for the cosmological constant?}
The standard lore on renormalization in Quantum Field Theory \cite{Collins} includes the fact that
the cosmological constant is additively renormalized away, the finite part being fully undetermined. There is an exception to this, however. In some cases, when nontrivial boundary conditions are imposed on the fields (like vanishing electromagnetic field in two fixed parallel plates in the classical example) the {\em difference} between the vacuum energy corresponding to nontrivial boundary conditions and the one with ``trivial" boundary conditions is computable and in many cases, finite. Confer \cite{Jaffe} for a clarifying review.    This effect is usually known as the Casimir effect and has nothing to do with the present work; boundary conditions are kept fixed (``trivial") in  our setting.
\par
 Let us review this in the presence of an external gravitational field. In order to do that, we shall consider a simple model of a scalar  quantum  field
\[
S=\int d^n x\sqrt{|g|}\left({1\over 2}g^{\m\n}\pd_\m\phi\pd_\n\phi-{\l\over 4!}\phi^4-{1\over 2}\xi R\phi^2+\a_3 R^2-\a_1 R^{\m\n\r\s}R_{\m\n\r\s}+\a_2 R^{\m\n}R_{\m\n}\right)
\]
(the terms quadratic in the curvature are necessary for renormalization). We do not believe that our  physical conclusions depend upon the details of the model.
Standard computation of the one-loop effective potential  \cite{Parker} yields the result
\[
V_{eff}=\Lambda_0+{1\over 2}\xi R\bp^2-\a_3 R^2-\a_1 R^{\m\n\r\s}R_{\m\n\r\s}-\a_2 R^{\m\n}R_{\m\n}+{1\over 2}m_0^2 \bp^2_0+{\l_0\over 4!}\bp^4_0+\nonumber
\]
\[
{\hbar\over 32 \pi^2}\left({1\over n-4}+{\g\over 2}-{3\over 4}\right)\left(m_0^2+{\l_0\over 2}\bp_0^2\right)^2+
{\hbar\over 64\pi^2}\left(m_0^2+{\l_0\over 2}\bp_0^2\right)^2\,log\,{m_0^2+{\l_0\over 2}\bp_0^2\over 4\pi \m^2}
\]
Defining renormalized quantities in the MS scheme
\[
\Lambda_0=\m^{4+\e}\left(\Lambda+{a_\Lambda(\l)\over n-4}\right)
\]
(where $\Lambda$ is dimensionless) leads to
\[
\b_\Lambda=-4\Lambda+{\hbar\over 32\pi^2}{m^4\over \m^4}
\]
The physical Cosmological Constant (CC) \footnote{
This is entirely analogous to  't Hooft's \cite{'tHooft} beta function for the dimensionless mass
\[
\b_m=-m+{\l m\over 32 \pi^2}
\]
in such a way that the physical mass 
\[
m_{phys}\equiv m \m
\]
and 
\[
\b_{m_{phys}}={\l m_{phys}\over 32\pi^2}
\]
}
is given by
\[
\Lambda_{phys}\equiv \m^4 \Lambda
\]
so that
\[
\b_{\Lambda_{phys}}={\hbar\over 32\pi^2}m^4
\]
Let us define the {\em cosmological constant}\footnote{Of course the CC is put by hand equal to zero if we were to use Coleman-Weinberg's \cite{Coleman} renormalization condition
\[
 V_{eff}(\bp=0)=0
\]  
}
  in the MS scheme as 
\[
\Lambda(\l)\equiv V_{eff}(\bp=0)
\]  

Assuming $\left.{\pd V_{eff}\over \pd \bp}\right|_{\bp=0}=0$, that is, absence of spontaneous symmetry breaking,
the CC obeys the renormalization group (RG) equation
\[
\left(\m{\pd\over\pd \m} +\b(\l){\pd\over\pd\l}+\b_\xi{\pd\over\pd\xi}+\sum_i \b_{\a_i}{\pd\over \pd\a_i}\right) \Lambda(\l)=0
\]
A successfully employed technique to compute the effective potential in many cases \cite{Einhorn} stems precisely from using this equation in a recursive way. Feeding the RG equation with the MS one loop result 
\[
\Lambda^1=\Lambda_0+{\hbar\over32\pi^2}\left({\g\over 2}-{3\over 4}\right)m^4+{\hbar\over 64\pi^2}m^4\,log\,{m^2\over \m^2}
\]
and 
using the well-known one-loop beta functions and anomalous dimensions \footnote{
To wit:
\bea
\b_{\Lambda_{phys}}={\hbar\over 32\pi^2}m^4 \textrm{ , }& \b_{m_{phys}}={\l m_{phys}\over 32\pi^2} \textrm{ , }& \b_\l={3\over 16\pi^2}\l^2\nonumber\\
\b_\xi={\l\over 16\pi^2}\left(\xi-{1\over 6}\right) \textrm{ , }& \b_{\a_1}=-{1\over 180}{1\over 16\pi^2} \textrm{ , }& \b_{\a_2}={1\over 180}{1\over 16\pi^2}\nonumber\\
\b_{\a_3}=-{1\over 72}{1\over 16\pi^2}{\xi\over 2}\left(\xi-{1\over 3}\right)&&
\eea
}
yields the $\m$ dependence of the two loop piece:
\[
\Lambda^2=\left({\l\hbar\over 256 \pi^4}\left({\g\over 2}-{3\over 4}\right)m^4+{\hbar\over 512\pi^4}\l m^4\,log\,{m^2\over 4\pi}+{\hbar\l m^4\over 1024 \pi^4}\right)log\,\m-\nonumber
\]
\[
{\hbar\over 512\pi^4}\l m^4\,\left(log\,\m\right)^2+f(g_i)
\]
where $g_i$ stands for {\em all} coupling constants not including the CC itself, but including the $\a_i$; this piece is not determined by the RG equations.
\par
No imaginary part appears ever by this procedure. Some ambiguity remains in the finite part independent of the coupling constants, which is usually fixed by the renormalization conditions, that is, renormalized to zero (\cite{Einhorn}). 
\par
The aim of the present work is to point out, following the logic of \cite{Polyakov} and the previous section that {\em generically} owing to lacking of translation invariance, the matter vacuum is unstable (that is, it can decay to physical particles with a certain computable width) so that by unitarity the vacuum energy must have a unambiguous finite imaginary part. Assuming analyticity in the coupling constants, this in turn puts restrictions on the possible real part of the cosmological constant through the dispersion relations as a consequence of Cauchy's theorem. This explicitly contradicts previous claims in the literature \cite{Dowker}.
\par
Another consequence of all this is that the usual analysis of quantum field theory in curved space-times is incomplete without a careful self-consistent consideration of the back-reaction problem, owing precisely to this imaginary part. We shall comment on this point in the conclusions.
\par

\newpage
\section{A model with unstable vacuum in flat space}
We have just witnessed a few paragraphs ago that explicit computations even in the simplest of background spacetimes, such as constant curvature  de Sitter or anti de Sitter spaces, are quite involved \cite{Bros}, and besides, the physical interpretation of the divergences found is not clear. 
\par
This is the rationale for first performing the complete analysis in a much simpler context in flat space, a {\em toy model} of sorts. It is clear that the main new ingredient of a curved space in this context is the non-conservation of four momentum. This can be achieved through an explicit time-dependent interaction term in flat space which violates energy-conservation. In the appendix the mapping between a subclass of scalar models in Friedmann spacetimes and flat models with spacetime dependent coupling constants is spelled out. 
\par

Let be the lagrangian
\be
\mathcal L=\frac12\partial\phi^2-\frac{m^2}2\phi^2-\frac{\lambda(t)}{3!}\phi^3
\ee
with a non-homogeneous coupling $\lambda(t)=\frac\lambda2\left(1+\cos\eta t\right)$. With this choice, energy is no longer conserved and a fixed amount ($\eta$) can be created or destroyed at any vertex. In the limit $\eta\to0$ we recover an standard $\lambda\phi^3$ theory.\\

With this interaction, the second order vacuum-to-vacuum amplitude reads:

\be
S=T\exp\left\{-\frac{i}{3!}\int\lambda(x^0):\phi(x)^3:dx\right\}
\ee

\begin{align}
\langle 0| S^{(2)}|0\rangle&=-\frac1{2\cdot3!}\int\lambda(x^0)\lambda(y^0)G(x,y)^3\,dxdy=\nonumber\\
&=i\frac{(2\pi)^3V\lambda^2}{8\cdot3!}\int dx^0dy^0\left\{\frac{dp_i}{(2\pi)^4}\right\}(1+\cos\eta x^0)(1+\cos\eta y^0)\frac{\delta^3\left(\mathbf p_T\right)e^{ip^0_T(x^0-y^0)}}{D_{p_1}D_{p_2}D_{p_3}}
\end{align}
where we abbreviate by $D_p$ the $p^2-m^2+i\epsilon$ denominators. $\{dp_i\}$ and $p_T$ mean $\prod_i dp_i$ and $\sum_i p_i$; $V$ is an infinite volume factor.\\

The time dependent coupling can be expressed as proportional to $1+\frac12 e^{i\eta x^0}+\frac12 e^{-i\eta x^0}$, so the ``energy conservation'' factors are now different
\begin{align}
\langle 0| S^{(2)}|0\rangle&=\nonumber\\
&=i\frac{(2\pi)^4VT\lambda^2}{8\cdot3!}\int \left\{\frac{dp_i}{(2\pi)^4}\right\}\frac{\delta^3\left(\mathbf p_T\right)}{D_{p_1}D_{p_2}D_{p_3}}\left(\delta(p_T^0)+\frac14(\delta(p_T^0-\eta)+\delta(p_T^0+\eta))\right)=\nonumber\\
&=i\frac{(2\pi)^4VT\lambda^2}{8\cdot3!}\left(T_{234}(0,m,m,m)+\frac12T_{234}(\eta^2,m,m,m)\right)
\end{align}
so the creation of energy at one vertex should be compensated in the other. $T$ is an infinite time factor. The standard integral $T_{234}$ is well known \cite{Berends} in general dimension:
\be
T_{234}(p^2,m_2,m_3,m_4)=\int \frac{d^nk_2 d^nk_3 d^nk_4}{(2\pi)^{3n}}\frac{\delta^n(k_1+k_2+k_3-p)}{(k_2^2-m_2^2+i\epsilon)(k_3^2-m_3^2+i\epsilon)(k_4^2-m_4^2+i\epsilon)}
\ee
and corresponds to the self energy ``setting sun'' diagram $\Sigma(\eta^2)$ in a $\lambda\phi^4$ theory.\\

Notice that we have assumed that $\eta\neq0$, so the regular $\lambda\phi^3$ theory result is not recovered when $\eta\to0$ in the last formula, i.e. the limit is discontinuous.\\

The corresponding vacuum decay rate can be calculated as the square of the T-matrix up to first order, $\mathcal T^{(1)}=-\frac1{3!}\int dx\lambda(x^0):\phi(x)^3:$
\begin{align}
\Gamma_{0\to3}&=\langle0|\mathcal T^{(1)}\mathcal T^{(1)\dagger}|0\rangle=\nonumber\\
&=\frac1{3!^2}\int dxdy\langle0|:\phi(x)^3::\phi(y)^3:|0\rangle=\frac1{3!}\int\lambda(x^0)\lambda(y^0)W(x,y)^3\,dxdy
\end{align}
We ca use now the expression for the Wightman function of the scalar field 
\be
\displaystyle W(x,y)=\int \frac{dp}{(2\pi)^4}(2\pi)\delta_+(p^2-m^2) e^{-ip(x-y)}=\int \frac{d\mathbf p}{(2\pi)^32E_p}e^{i\mathbf p(\mathbf x-\mathbf y)}e^{-iE_p(x^0-y^0)}
\ee
and then
\begin{align}
\Gamma_{0\to3}&=\frac{(2\pi)^3V\lambda^2}{3!4}\int dx^0dy^0\left\{\frac{(2\pi)\delta_+(p^2_i-m^2)dp_i}{(2\pi)^4}\right\}(1+\cos\eta x^0)(1+\cos\eta y^0)e^{-ip^0_T(x^0-y^0)}\delta^3(\mathbf p_T)=\nonumber\\
&=\frac{(2\pi)^4VT\lambda^2}{3!4}\int \frac{(2\pi)\delta_+(p^2_i-m^2)dp_i}{(2\pi)^4}\left(\delta(p_T^0)+\frac14(\delta(p_T^0-\eta)+\delta(p_T^0+\eta))\right)\delta^3(\mathbf p_T)=\nonumber\\
&=\frac{(2\pi)^4VT\lambda^2}{3!16}\int \frac{(2\pi)\delta_+(p^2_i-m^2)dp_i}{(2\pi)^4}\delta(p_T^0-\eta)\delta^3(\mathbf p_T)=\nonumber\\
&=\frac{(2\pi)^4VT\lambda^2}{3!16}\int \frac{d\mathbf p_1d\mathbf p_2}{(2\pi)^98E_1E_2E_{12}}\delta(E_1+E_2+E_{12}-\eta)
\end{align}
where $E_{12}=\sqrt{(\mathbf p_1+\mathbf p_2)^2+m^2}$. From the three delta function factors for te energy, only the one with positive increment of energy contributes, for kinematical reasons.\\

This expression is just the standard three-body phase space factor
\be
\Gamma_{0\to3}=\frac{(2\pi)^4VT\lambda^2}{3!16}\int \frac{dE_1dE_2}{4(2\pi)^7}\theta(E_T^\text{max}-\eta)\theta(\eta-E^\text{min}_T)
\ee
where $E^{{\vphantom{a}^\text{max}_\text{min}}}=E_1+E_2+\sqrt{(p_1\pm p_2)^2+m^2}$ correspond to parallel and anti-parallel configurations of momenta. There is also an implicit kinematic factor $\theta(\eta-3m)$ indicating the minimun energy necessary to create 3 particles.\\

The final expression for this rate is
\be
\Gamma_{0\to3}=\frac{(2\pi)^4VT\lambda^2}{3!16}\int_m^\frac{\eta^2-3m^2}{2\eta}\frac{dE}{4(2\pi)^7}I(E,\eta)
\ee
with
\be
I(E,\eta)=\frac{\sqrt{(E^2-m^2)(\eta^2+m^2-2E\eta)(\eta^2-3m^2-2\eta E)}}{m^2+\eta^2-2\eta E}
\ee
This rate is proportional to the decay rate of a particle with mass $\eta$, $\Gamma_{M=\eta}$. We can give he approximate result of this integral if $\eta$ is very close to or much bigger than $3m$:
\be
\int_m^\frac{\eta^2-3m^2}{2\eta} dE I(E,\eta)\simeq 2\sqrt{3}m^2\epsilon^2\text{ , if  }\ \eta=3m(1+\epsilon)
\ee
\be
\int_m^\frac{\eta^2-3m^2}{2\eta} dE I(E,\eta)\simeq\frac{\eta^2}8\text{ , if }\ \eta\gg m
\ee\\

From the application of Cutkosky's rules, the ``setting sun'' self-energy diagram has an imaginary part given by the corresponding decay rate, $\text{Im}\,\Sigma(\eta^2)=\Gamma_{M=\eta}$, which is based in the identity \cite{Berends}
\be
\int_m^\frac{\eta^2-3m^2}{2\eta}dE\,I(E,\eta)=\frac1{4\pi^5}\text{Im}\left((2\pi)^{3n} T_{234}\right)^\text{fin}\Big|_{n=4}
\ee

Since the processes in our model are directly related to the aforementioned diagrams, we can establish easily the unitarity relation
\be\label{basic}
-2\text{Re}\langle 0|S^{(2)}|0\rangle=\Gamma_{0\to3}
\ee

In the previous calculation, we avoided to deal with the tadpole diagrams by taking the interaction operator to be normal ordered. If we choose to include those extra diagrams, the unitarity relation still holds. The tadpole contribution in (\ref{basic}) is then:
\be
-2\text{Re}\langle 0|S^{(2)}|0\rangle=\Gamma_{0\to3}+\Gamma_{0\to1}
\ee

Here the left member acquires a new term which is
\be
\langle 0|S^{(2)}_\text{tad}|0\rangle=-\frac18\int dx dy \lambda(x^0)\lambda(y^0)G(x,x)G(y,y)G(x,y)
\ee
The closed loops $G(x,x)$ and $G(y,y)$ are constants, and they are always real (at least in dimensional regularization) and so, not very relevant for the calculation.
\begin{align}
\langle 0|S^{(2)}_\text{tad}|0\rangle&=-\frac i{8\cdot16}G_0^2\int \frac{dx dydp}{(2\pi)^4} (2+2\cos\eta x^0)(2+2\cos\eta y^0)\frac{e^{ip(x-y)}}{D_p}=\nonumber\\
&=-\frac {i (2\pi)^2 V_3}{8\cdot16}G_0^2\int dx^0 dy^0dp^0 (2+2\cos\eta x^0)(2+2\cos\eta y^0)\frac{e^{ip^0(x^0-y^0)}}{(p^0)^2-m^2+i\epsilon}=\nonumber\\
&=-\frac {i (2\pi)^4 V_4}{8\cdot16}G_0^2\left(\frac4{m^2}+\frac{2}{\eta^2-m^2+i\epsilon}\right)
\end{align}

In the other hand, the total decay rate $\Gamma_{0\to1}$ is 
\begin{align}
\Gamma_{0\to1}&=\frac1{3!^2}\sum_{1-\text{particle}}\int dxdy\lambda(x^0)\lambda(y^0)\langle0|\phi(x)^3|1\rangle\langle1|\phi(y)^3|0\rangle=\frac14G_0^2\int dxdy\lambda(x^0)\lambda(y^0)W(x,y)=\nonumber\\
&=\frac1{64}G_0^2\int\frac{dxdyd\mathbf p}{(2\pi)^32E_\mathbf{p}}e^{i\mathbf p(\mathbf x-\mathbf y)}(2+2\cos\eta x^0)(2+2\cos\eta y^0)e^{-iE_\mathbf{p}(x^0-y^0)}=\nonumber\\
&=\frac{(2\pi)^5V_4}{64}G_0^2\frac1{2m}\left(\delta(\eta-m)+\delta(\eta+m)\right)
\end{align}
We have assumed $W(x,x)=G(x,x)$.\\

These new contributions to (\ref{basic}) should match each other. It is easy to check that they actually do by applying the Weierstrass theorem
\be
\frac i{\eta^2-m^2+i\epsilon}=\frac{\pi}{2m}\left(\delta(\eta-m)+\delta(\eta+m)\right)+\ldots
\ee
where the dots indicate imaginary contributions. No infinities nor ambiguities have been encountered in our toy model.
\newpage
\section{Kinetic equations}

The effects shown in the previous section, allowed because of the non-conservation of energy, led to the production of particles in the initially empty space. In our very simplified model there are no conserved quantum numbers; but in general the set of produced particles must enjoy the quantum numbers (id est, charges) of the vacuum. An important physical question is towards which final state this instability leads to? Again, this question has many faces. In the general case in which a gravitational external field is present, the backreaction is surely important, but difficult to compute. It is not even clear that the usual procedure of solving again Einstein's equations with a second member given by the expectation value of some energy-momentum tensor \cite{PerezNadal} would be  good enough for our purposes.
\par
A second facet of this problem is the evolution of the particle (in general, conserved quantities) density.
 We can study this phenomenon by considering a spacetime box $V\times T$, with an initial number of particles (in our simplified model there is only one type of particles) $N$, do that the initial density is $n\equiv {N\over V}$ \footnote{
 A finer description would be provided by the Wigner function, in which both momentum and position distributions are correlated, to the extent that this is compatible with Heisenberg's principle. This construct obeys suitable generalizations of Boltzmann's equation.}

It is well-known \cite{Weinberg} that the transition {\em rate} $d\Gamma\left(\a\rightarrow \b\right)\equiv {d P\left(\a\rightarrow \b\right)\over T}$ of a given process in which $N_\a$ particles in the initial state evolve into $N_\b$ particles in the final state depends on the three-space volume, $V$,  
as $V^{1-N_\a}$. This clearly means that the vacuum decay terms in  Boltzmann's equation (or whatever improvement thereof) will clearly dominate, because they are the only ones which are {\em extensive}, id est, proportional to the ordinary volume.
Let us build up a simplified model for this equation. The essential thing is to capture the volume as well as the power of the distribution function itself. Taking into account the vacuum decay and absorption only,  the balance  equation in our fiducial spacetime box is of the schematic form
\[
{dN_{\mathbf p}\over dt}=\Gamma^{\mathbf p}_{03}-\Gamma^{\mathbf p}_{30} 
\]  
 Here $N_{\mathbf p}\, d\mathbf p $ represents the total number of particles with momentum between $\mathbf p$ and $\mathbf p+ d\mathbf p$ in the fiducial volume $V$, and $T$ is our fiducial time. The constructs $\Gamma^{\mathbf p}_{03}$ and $\Gamma^{\mathbf p}_{30}$ denote the amplitude for vacuum decay or annihilation to the vacuum when precisely one of the particles has momentum $\mathbf p$.  The transition rate is simply related to the transition probability $dP(\a\rightarrow\b)$ by $ d\Gamma(\a\rightarrow\b)={dP(\a\rightarrow\b)\over T}$. General arguments implicate
\[
d\Gamma(\a\rightarrow\b)=(2\pi)^{3N_\a-2} V^{1-N_\a}|M_{\b\a}|^2\d^4(p_\a-p_\b)d\b
\]
where the matrix $M$ is intimately related to the matrix $S$:
\[
S_{\b\a}\equiv -2\pi i \d^4(p_\a-p_\b) M_{\b\a}
\]
and 
\[
d\b\equiv \prod_{i\in N_\b} d\mathbf p_i
\]
With this proviso, the mass dimension of $|M|^2$ is $8-3(N_\a+N_\b)$.
\par
It is now clear that the increase in the density of particles of momentum $\mathbf p$ out of the vacuum in the fiducial volume $V$ during the time interval $T$ is
\[
\Gamma_{03}^{\mathbf p}\equiv \int d\Gamma_{03}(\mathbf p,\mathbf p^\prime_2,\mathbf p^\prime_3)\,d\mathbf p^\prime_2\,d\mathbf p^\prime_3 (1+N_{\mathbf p})(1+N_{\mathbf p_2})(1+N_{\mathbf p_3}).
\] 
The decrease in the density of  particles of momentum $\mathbf p$ due to annihilation  to the vacuum (id est, the reverse process) in the same fiducial volume and time interval is proportional to the existing density cube of particles:
\[
\Gamma_{30}^{\mathbf p}\equiv \int N_{\mathbf p} N_{\mathbf p_2} N_{\mathbf p_3}\,d\mathbf p_2\,d\mathbf p_3 \,d\Gamma_{30}\left(\mathbf p,\mathbf p_2,\mathbf p_3\right)
\]
The total (integrated) widths are given by 
\be
\Gamma_{03}=\frac{V\lambda^2}{3!64 (2\pi)^3}\int_m^\frac{\eta^2-3m^2}{2\eta}dE\,I(E,\eta)\simeq \frac{V\lambda^2}{\sqrt{3}\,64 (2\pi)^3}m^2\epsilon^2
\ee
where the last equality holds approximately when $\eta=3m(1+\e)$.
\begin{align}
\Gamma_{30}&=\int d\mathbf p_1d\mathbf p_2d\mathbf p_3 N_{\mathbf p_1}N_{\mathbf p_2}N_{\mathbf p_3} d\Gamma_{30}(\mathbf p_1,\mathbf p_2,\mathbf p_3)=\nonumber\\
&=\frac{\lambda^2(2\pi)^4}{16 V^2}\int N_{\mathbf p_1}N_{\mathbf p_2}N_{\mathbf p_3}\frac{d\mathbf p_1}{2E_1}\frac{d\mathbf p_2}{2E_2}\frac{d\mathbf p_3}{2E_3}\delta^3(\mathbf p_1+\mathbf p_2+\mathbf p_3)
\end{align}

A toy model that embodies some of the characteristics of our integro-differential equation is

\[
{dn\over dt}=C\left(M^4- {n^3\over M^5}\right) 
\]
where $C$ is a constant and we have saturated all dimensions with an average mass scale, $M$.
It is clear that in any such model the density of particles will rapidly grow as
\[
n\sim C M^4 t
\]

until it becomes so big that
\[
n\sim M^3
\]
which happens in a characteristic time
\[
\t\equiv {1\over C M}
\]
 
\newpage
 \section{Conclusions}
  In this paper the issue of vacuum instability in the presence of a cosmological constant has been discussed. Some computations for conformally invariant matter in de Sitter space were done, and an ambiguity with Fermi's Golden Rule discussed in some detail.
 \par
 A flat space model with time-dependent coupling constant, that captures some of the aspects of the physically interesting situation has been introduced, and analyzed to one-loop order. No ambiguities have been encountered in this piece of work.
 \par
It is somewhat hazardous to transport all lessons from this model to the quantum fields interacting in an external gravitational background, but it is clear at any rate that vacuum  instability is a generic phenomenon for any quantum field in the presence of such an external gravitational background. Any attempt to forbid it by using clever quantum numbers seems doomed by the fact that it would always be possible to create particle-antiparticle pairs with vacuum quantum numbers. Now there are not so many logical possibilities.
\par
One  that immediately comes to mind is that {\em unitarity} can be violated. We have pointed out already in section \ref{vacdecay} that what is essential here is not the full fledged notion of unitarity, but only the milder consequence of the definition of Feynman's propagator that is usually known as the {\em largest time equation}. In flat spacetime this is known to implicate unitarity, but perhaps in a general background this need not be so; nevertheless the largest time equation itself seems unavoidable as long as Feynman's propagators are used. It could be that ordinary perturbation theory gets modified in this context; careful computations \cite{Bros} seem to indicate that this is not the case, although the final word has not been said .
\par
If unitarity in this sense holds true, then, as we have argued, the cosmological constant gets a nonvanishing imaginary part, which points to the existence of some instability. Now what could be the endpoint of such an instability? It is clear that the {\em only} spacetime in which this instability is absent is flat space-time.

\par
 It seems difficult at any rate to resolve this issue in the hybrid framework \cite{Barvinsky} in which the quantum expectation value of the energy-momentum tensor is used as a source for Einstein's equations. It is likely that the gravitational field has to be treated in equal terms with all other fields in order for a  self-consistent solution to exist.
\par
Detailed calculations in this framework are necessary but difficult. We hope to report on them in the near future.
\section*{Acknowledgements}
We are grateful to Andrei Barvinski, Joaqu\'{\i}n D\'{\i}az-Alonso, Jaume Garriga and Enric Verdaguer  for illuminating discussions. We  also thank Atsushi Higuchi, Ugo Moschella and Alexander Polyakov for useful correspondence.
This work has been partially supported by the
European Commission (HPRN-CT-200-00148) as well as by FPA2009-09017 (DGI del MCyT, Spain) and 
 S2009ESP-1473 (CA Madrid). R.V. is supported by a MEC grant, AP2006-01876.

\newpage
\appendix
\section{Alternative computation of the vacuum decay rate.}
Let us review the  calculation in \cite{Higuchi}.  In the conformally flat patch 
\[
ds^2={l\over u}^2(du^2-d\mathbf x^2)
\]
 for a conformally coupled scalar field
\footnote{A conformally coupled scalar field is massless and the value of the curvature coupling is $\xi=-\frac14\frac{n-2}{n-1}$, where $n$ is the dimension of spacetime. In de Sitter space, where the curvature is constant, $R=-{n(n-1)\over l^2}$, this is equivalent to a minimally coupled ($\xi=0$) scalar field with mass $m^2=-\frac14\frac{n-2}{n-1} R={n(n-2)\over 4l^2}$. This mass is in the complementary series, with a $i\mu=\frac12$ parameter ($m^2 l^2=\mu^2+{(n-1)^2\over4}$) , and its corresponding two-point functions are particularly simple $F\left(\frac n2,\frac n2-1,\frac n2;\frac{1+z}2\right)\propto (1-z)^{1-\frac n2}$.}
we can expand the free field expansion in term of the modes (we represent by $k\equiv |\mathbf{k}|$).
\[
f_\mathbf{k}(u,\mathbf x)=\frac{H u}{(2\pi)^3\sqrt{2k}}e^{iku}e^{i\mathbf{kx}}
\]
which are equivalent to choosing the Euclidean or Bunch-Davies vacuum\footnote{For the n-dimensional case, the appropriate modes are $f_\mathbf{k}(u,\mathbf x)=\frac{(H u)^{\frac n2-1}}{(2\pi)^{n-1}\sqrt{2k}}e^{iku}e^{i\mathbf{kx}}$}. The normalization is such that $[a_1,a_2^\dagger]=(2\pi)^3\delta_{12}$. This expansion is the same for the interacting field in the interaction picture, so the amplitude $0\to\mathbf{k}_1\mathbf{k}_2\mathbf{k}_3\mathbf{k}_4$  is simply
\begin{align}
M(\mathbf{k}_i)&=\int_0^\infty\,du\, d\mathbf{x}\langle\mathbf{k}_1 \mathbf{k}_2 \mathbf{k}_3 \mathbf{k}_4|\frac\lambda{4!}\phi(u,\mathbf{x})|0\rangle=\lambda(2\pi)^{12}\int\,du\,d\mathbf{x} \prod_{i=1}^4f^*_{\mathbf{k}_i}(u,\mathbf{x})=\nonumber\\
&=\lambda\int\,\frac{du d\mathbf x}{4\sqrt{k_1k_2k_3k_4}}e^{-iu\sum_i k_i}e^{-i\sum_i \mathbf{x}.\mathbf{k}_i}=\lambda(2\pi)^3\int\,\frac{du }{4\sqrt{k_1k_2k_3k_4}}e^{-iu\sum_i k_i}\d(\sum_i\mathbf{k}_i)
\end{align}
The total decay rate
\[
P=\frac1{4!}\int\,\prod_{i=1}^4\frac{d\mathbf k_i}{(2\pi)^3}|M(\mathbf{k}_1,\mathbf{k}_2,\mathbf{k}_3,\mathbf{k}_4)|^2=\frac{\lambda^2(2\pi)^3V_c}{4!(2\pi)^{3\cdot 4}16}\int\,dudv\prod_{i=1}^4\frac{d\mathbf k_i}{k_i}e^{-i(u-v)\sum_i k_i}\delta(\mathbf \sum_i \vec{k}_i)
\]
We consider that the infinite quantity 
\[
V_c=(2\pi)^3\delta(\mathbf 0)
\]
 represents the volume of 3-space. The following changes of variables are considered:
\bea
 &u\equiv l\,e^{-{t_1\over l}},\ &2T=t_1+t_2\nonumber\\
  &v\equiv l\,e^{-{t_2\over l}},\ &\tau=t_1-t_2;
 \eea
  yielding
\[
P=\frac{\lambda^2 V_c}{4!(2\pi)^{9}16}\int_{-\infty}^\infty\,dT\,d\tau\prod_{i=1}^4\frac{d\mathbf k_i}{k_i}\delta(\sum_{i=1}^4\mathbf{k}_i)
e^{-2{T\over l}}\exp\left\{2il\sum_i k_i\,e^{-HT}\sinh{\tau\over 2l }\right\}\]

Now the integral over the momenta includes a term which looks singular
\[
\int d\mathbf k_1 d\mathbf k_2 d\mathbf k_3 {1\over k_1 k_2 k_3 |\mathbf{k}_1+\mathbf{k}_2+\mathbf{k}_3|}
\]
The point is that $|\mathbf{k}_1+\mathbf{k}_2+\mathbf{k}_3|$ can well vanish even if all  $k_i\neq 0$.
 Expanding around the seemingly singular region, it appears that the singularity is integrable.
 
 \par
 
At any rate, the integral diverges for high momenta, unless the exponential decreases, which happens only when
\[
\text{Im} \sum_i k_i \geq 0
\] 

Nevertheless, it can be defined as a Gamma function by analytic continuation, so that let us carry on.


In the reference we are annotating  unity is introduced in the form 
\[
\int_0^\infty\,dK\delta(\sum_{i=1}^4 k_i-K)=1
\]
 and  the momenta are normalized by $\mathbf k_i=K\mathbf y_i$, so that using 
 \[
 \int\prod_{i=1}^4\frac{d\mathbf y_i}{y_i}\delta(\sum_i y_i-1)\delta(\sum_i\mathbf{y}_i)=\frac{\pi^3}{4}
 \]
 yields
\[
P=\frac{\lambda^2 V_c}{3(8\pi)^{6}}\int\,dTd\tau dK K^4 e^{-2{T\over l}}\exp\left\{\frac{2iK}He^{-{T\over l}}\sinh{\frac{\tau}2 l}\right\}=
\]
\[
=\int dT e^{3{T\over l}} V_c \left(\frac{\lambda^2}{48 l^4(8\pi)^6}\int_0^\infty d\kappa\kappa^4\int d\eta\exp(i\kappa\sinh\eta)\right)
\]

where $\kappa=2K l e^{-{T\over l}}$ and $\eta={\tau\over 2 l}$ the particle production rate per unit volume is
\[
\Gamma=\frac{\lambda^2}{48 l^4 (8\pi)^6}\int_0^\infty d\kappa\kappa^4\int d\eta\exp(i\kappa\sinh\eta)=\frac{\lambda^2}{48 l^4 (8\pi)^6}\int_0^\infty d\kappa\kappa^42K_0(\kappa)=\frac{3\lambda^2 }{4 l^4 (16\pi)^5}
\]
This formula is very appealing physically; it gives a finite result for a tree-level cross section, which is a good thing, because the only known place in which divergent cross sections appear at tree level is in bremsstrahlung effect (correction to external legs by emission of massless particles), and this is not our case.
\footnote{In the last formula the integral representation for the Bessel function $K_0(z)$ has been used. This is for $-\pi/2\leq Arg\,z\leq \pi/2$

\[
K_0(z)=\int_0^\infty e^{-z\,cosh\,t} dt=\int_0^\infty e^{iz\,sinh\,(t+i\pi/2)} dt=\int_{i\pi/2}^{\infty+i\pi/2} e^{iz\,sinh\,t} dt
\]
which seems to require that $Im\,\eta=\pi/2$, which is not the case, but again the integral can be interpreted as the analytical continuation of a convergent one.
}

\par 
This means that whereas the calculation in section 2 was proportional to the covariant spacetime volume, $V_4$ (times a divergent expression), Higuchi's is proportional to $V_c$ times the integral over $dT e^{3{T\over l}}$ times a {\em finite} expression. If we identify
\[
V_4\sim V_3\int dT e^{3{T\over l}}
\]
both calculations are inconsistent. But this identification is not compulsory. As has been already pointed out in \cite{Bros} it is not clear in which coordinates the particle production per unit volume per unit time is to be defined.
\par
Ay any rate, by consistency, in case the computation of \cite{Higuchi} is preferred, the imaginary part of the free energy should also be finite.
\section{Relationship between time-dependent coupling constant and cosmological backgrounds}
Let us look for the cases in which a Minkowski space lagrangian with time-dependent coupling constants such as
\[
L={1\over 2} \pd_\m \psi\pd^\m\psi-{m(t)^2\over 2}\psi^2-{\l(t)\over 6}\psi^3
\]
is equivalent to a scalar field in a curved, conformally flat gravitational background
\[
ds^2=a(t)^2 \eta_{\m\n}dx^\m dx^\n
\]
namely
\[
L=\sqrt{\|g\|}\left({1\over 2}g^{\m\n}\pd_\m \phi \pd_\n\phi-{m^2\over 2}\phi^2-{\l\over 6}\phi^3\right)={a^2\over 2}\left(\dot{\phi}^2-(\phi^\prime)^2\right)-a^4{m^2\over 2}\phi^2-a^4{\l\over 6}\phi^3
\]

This is always possible, up to a total derivative, with the identifications
\bea
&&\psi=a\phi\nonumber\\
&&\l(t)\equiv \l a(t)\nonumber\\
&&m^2(t)\equiv m^2 a^2-{\ddot{a}\over a}
\eea

It would also be interesting to find a mapping for the different vacuum states as well.

\newpage 

\end{document}